# Enhance your smartphone with a Bluetooth Arduino nano board

F. Bouquet[1], G. Creutzer[1], D. Dorsel[2], J. Vince[3] and J. Bobroff[1]

[1] Université Paris-Saclay, CNRS, Laboratoire de Physique des Solides, 91405, Orsay, France.
[2] Institute of Physics I and II, RWTH Aachen University, 52062 Aachen, Germany.
[3] IFé, ENS Lyon, ESPE Lyon, 15 parvis René Descartes, BP 7000, 69342 Lyon Cedex 07 France.

**Abstract**

Using smartphones in experimental physics teachings offers many advantages in term of engagement, pedagogy and flexibility. But it presents the drawbacks of possibly endangering the device and also facing the heterogeneity of available sensors on different smartphones. We present a low-cost alternative that preserves the advantages of smartphones: using a microcontroller equipped with a large variety of sensors that transmits data to a smartphone using Bluetooth Low-Energy protocol. This device can be lent to students with little risks and used to perform a wide range of experiments. It opens the way to new types of physics teachings.

## 1. Introduction

Using smartphone in physics classes is more and more common, as demonstrated by the growing literature (see e.g. [Hochberg2018][Staacks2018][Chevrier2013]). It presents many advantages: familiarity of the students with the device, possibility of remote experimental activities, new pedagogical approaches, and lately lockdown experimental activities [Obrien2020]. However, specific issues often prevent the use of smartphones in teaching activities:

- there can be a risk of damaging a student's smartphone in some experiments;
- the sensors that are available will vary a lot depending on the smartphone, making it difficult for a teacher to plan a common class activity (for example barometers are not common sensor on smartphones; the iPhone light sensor data are not available, …);
- some sensors essential to physics are not present in any smartphones (such as thermometer or hygrometer), restricting the possible activities.

We present here a low-cost solution that help solving these three issues and increases the flexibility offered by using a smartphone as a visualization and measuring device.

## 2. Setup

The setup consists in using an Arduino nano 33 BLE sense microcontroller [nano] as a measuring tool (see Figure 1). This board comes with many sensors directly mounted on the board: accelerometer, gyroscope, light sensor, temperature and hygrometer, barometer, magnetometer, microphone, plus of course an analog-to-digital converter. The originality of this board is that it also contains a Bluetooth Low-Energy (BLE) module, making it possible to wirelessly connect the board to a smartphone.

To do so, we used a Bluetooth library for connecting external devices to the phyphox application. [Staacks2018] The opensource phyphox application is well known among physics teachers as one of the apps that can access the smartphone sensor data. One of its particularity is that it offers the possibility of coding your own interface to customize your experiment (data processing, plotting, saving, …) and ways of easily share it with others. To use the nano board, we developed some simple phyphox programs that, once install on a smartphone:

- Enable the BLE connection;
- Send to the nano board an ID number that characterizes the experiment that is expected;

- Wait for data sent from the nano board, and plot them on the smartphone screen when they arrive.

In parallel, we developed a program for the nano board that:
- Installs all the libraries for BLE communication and sensor utilisation;
- Waits for an experiment ID;
- Upon receiving the experiment ID, starts the relevant sensors and send the expected data to the smartphone.

None of these programs are very complicated and they can easily be adapted to local conditions and needs. They are open-source (free to use and modify), and available as supplementary digital materials to this article, as well as a comprehensive instruction set.

In practice, the nano board needs to be loaded with its program only once, and then becomes an autonomous device that can stored, and used with minimum setup time when needed. The nano board is very flexible, since its size is small (45 × 18 mm) and it can be easily powered by batteries or a USB charger. USB power banks can also be used provided they do not have an auto switch-off feature. For our usage we have developed a 3D-printed case with space for two button cells, making a very compact object that the students can easily use in various settings. The phyphox experiments are distributed to students using QRcodes, a quick and easy method that reduces the setup time to minimum.

### 3. Teaching using a nano board

The variety of available sensors on the nano board, plus the possibility of using any Arduino compatible external sensor, open a wide range of possible experiments: mechanics, acoustics, optics, … The physics usage examples we propose hereafter are based on our own teaching experience with students in our University or in high schools in France. Such teachings were developed thanks to the fact that the use of microcontrollers and smartphones have been encouraged in French national curricula recently.

*Students' laboratory classes*

The nano board can be used during a normal students' laboratory class as a measuring device (at high school or university level). The nano board does not require coping with breadboards and electronic circuit building, and can be connected rapidly with student's smartphone. It is small enough to be used in various setups. For example, in classical mechanics, it allows to measure the timing of the free fall of a body, the period and damping of a simple pendulum, a friction coefficient on a tilted plate…

*Classroom demonstration*

Classroom demonstrations show a physics phenomenon to students in a very concrete way, and can be used to bring rhythm into a lecture. Wireless data monitoring opens up new possibilities: by casting onto a video projector the screen of the professor's smartphone connected to a nano board, it is possible to let a whole class monitors in real time the results of an experiment, which makes visible a quantity that is not easy to grasp.

For example, when studying the free fall, displaying the acceleration of an object thrown upward, doing the experiment and looking at the data at the same time, would address this typical students' misconception [Clement1982]. The pressure inside a vessel connected to a water pump can be measured and correlated to the flow of water. Another example is to investigate the variation of pressure inside a rubber balloon when it is inflated. The nano board can fit inside a balloon and the inside pressure can be recorded. The light intensity measured by the nano board can be used as an indication of the rubber thinness. Observing the different regimes of the pressure inside the balloon [Stein1958], all the while expecting it to burst, makes for a climatic demonstration for students.

*At-Home Experiments*

Having students performing at-home experiments is possible now that smartphones are ubiquitous. At-home experiments can be proposed as a pedagogical exercise [bouquet2018], but also imposed by a pandemic situation [obrien2020]. Lending a nano board to students increases the number of possible experiments, either by providing new sensors (such as a thermometer, for thermodynamic experiments), or by providing a substitute for the students' smartphones, avoiding the risk of damaging them (which can be a risk in some mechanical experiments, such any free fall experiments requiring to throw a smartphone in the air).

For example, students can be asked to study simple thermodynamics. Enclosing the nano board into an airtight vessel (emptied jar of jam), putting it into the freezer, and monitoring both the temperature and pressure allows to test the variation of pressure when the temperature is changing from room to freezing temperature (see figure 2). Qualitative analysis works well, precise quantitative analysis requires some care since the thermometer thermal inertia is larger than the barometer response time. Students can also experimentally verify that the light inside a fridge indeed goes off when the door is closed (measuring the pressure – to detect the opening and closing of the door – and light intensity in parallel).

Another example is to study the thermal relaxation of a body of water. After wrapping the arduino nano into a watertight plastic bag, it can be immerged into a volume of hot water. Assuming simple heat-loss model, an exponential relaxation is expected for the temperature (see figure 3), and heat-loss values can be extracted and compared to housing isolation values [bouquet2018]. An alternative approach would be to use an Arduino-compatible waterproof thermometer as an external sensor and connect it to the nano board (the data would still be received on the smartphone).

*Physics projects*

Letting students work on open physics projects allows them explore experimental physics outside the controlled framework of a students' laboratory [Bouquet2017]. Since the nano board is not expensive, it can easily be lent to students for projects that span over several weeks.

For example, using a nano board, third-year university-level physics students could perform the Beer-Lambert experiment [bouquet2018][Onorato2018] in autonomy, outside of the classroom, as part of an experimental project class. The nano board allowed students owning an iPhone to participate (the iPhone light sensor access is blocked to users), and moreover the nano board light sensor can measure the light intensity on three different wavelengths (red, green, blue), allowing for some crude spectrometry experiments (see figure 3).

*Coding and electronic activities*

We used the nano board as a "black box" for our students, during experimental physics-oriented classes. But as any Arduino board, it can be used as a nice playground to teach coding or electronics to students. The easy connection to a smartphone allows for a quick feedback of what is measured by the board and can be used to help understanding the measuring process and the limits of the sensors. The fact that all the programmes are open-source also allows for the most computer-oriented students to deep further into advanced coding.

## Conclusion

Using an Arduino nano board offers a cheap (less than 40 $) solution to solve some problems raised by the use of smartphone to do a physics experiment, and could help physics teachers to develop new activities for their students. It is easy to adapt this setup to different sensors or experiments, and it increases the flexibility of the use of smartphone. It also opens the possibility of experimental homework or to engage students in project-based activities.

## Acknowledgements

We gratefully acknowledge Sebastian Staacks and Jens Noritzsch for discussions and inspiration. We thank the students who participated to these projects, and the Institut Villebon - Georges Charpak for helping us to develop new teachings. This work has been partially supported by the Chair "Physics Reimagined" led by Paris-Saclay University and sponsored by AIR LIQUIDE.

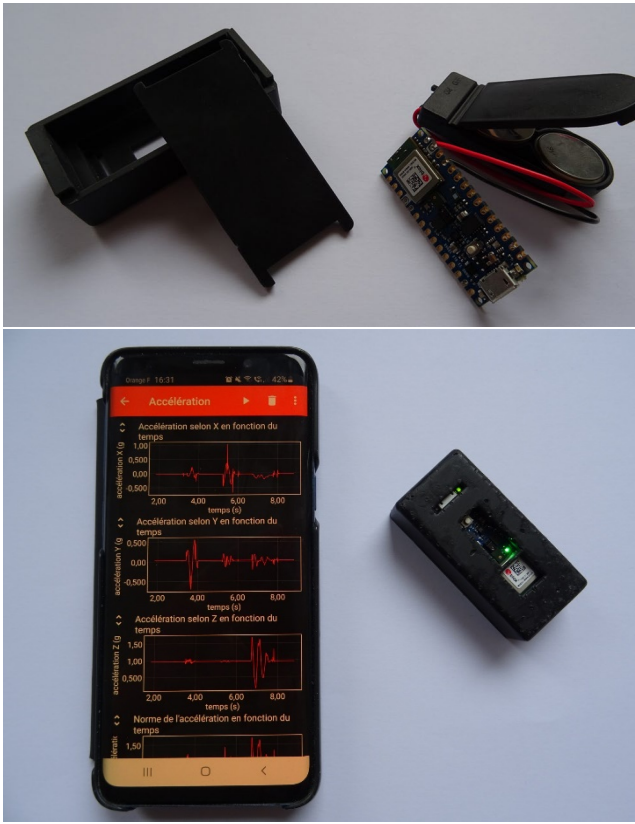

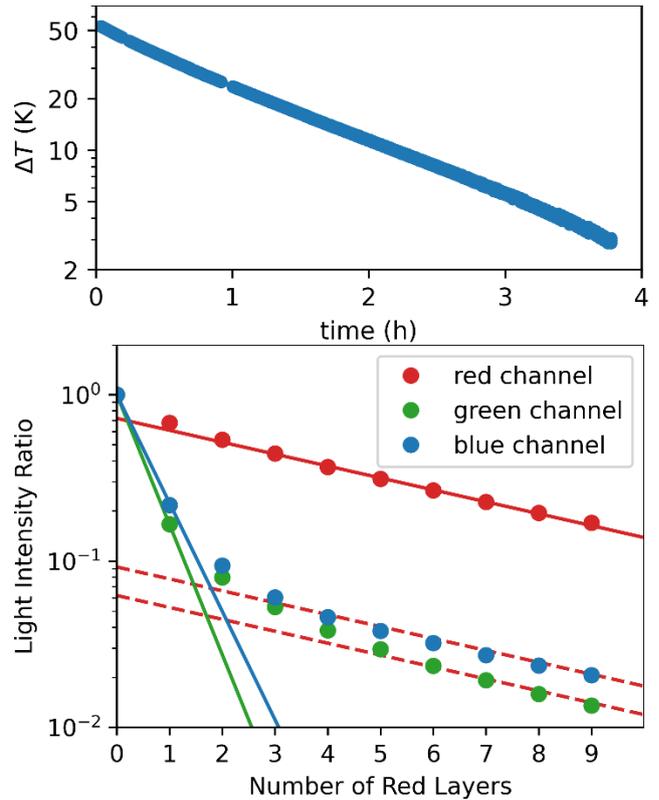

Figure 1: Top: the arduino nano bord connected to two button cells (right), next to a 3D printed case (left). Bottom: the arduino board and the button cells are in the case, transfering data to the smartphone via Bluetooth Low-Energy.

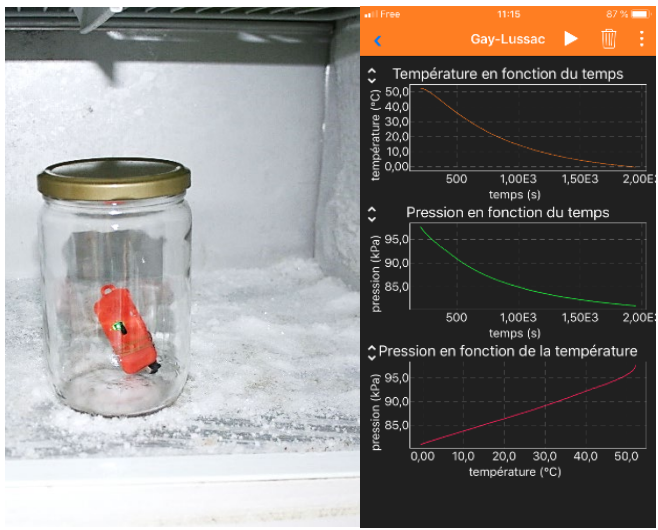

Figure 2: Left: experimental setup for the ideal gas law experiment: the nano board is enclosed into an airtight jam jar. It was first heated up to 50°C into and oven, then put into a freezer. Right: screen of the smartphone monitoring the experiment.

Figure 3: Data obtained with the nano board. Top: temperature of a nano board immerged into of a vessel of hot water (The ambient temperature has been subtracted to the temperature of the water). The linear behaviour shows an exponential thermal relaxation. Bottom: ratio of light intensity that passes through layers of red transparent sheets, in function of the number of sheets, for the red, green and blue channels of the light sensor. An exponential behaviour (Beer-Lambert law) gives a linear behaviour in this representation. The red transparent sheets are letting the red wavelength mainly go through; the red line indicates that only 15% of the light intensity is lost per sheet. The blue and green wavelengths are blocked by just a couple or red sheets; when more layers are added, the variation of the signals measured on these two channels, parallel to the red channel measurements, shows that a small fraction of the red wavelength is recorded on these two channels. Similarly, the fact that the linear fit of the red channel data intersects the axis for zero layer at 72% and not 100% shows that a fraction of the green and blue wavelengths is also recorded on that channel.

4